\documentclass{article}
\usepackage{amsmath,graphicx,mlspconf}

\usepackage{xcolor}
\usepackage{multirow} 
\usepackage{float}
\usepackage{hyperref}

\copyrightnotice{978-1-6654-8547-0/22//\$31.00 {\copyright}2022 IEEE}

\toappear{2022 IEEE International Workshop on Machine Learning for Signal Processing, Aug.\ 22--25, 2022, Xi'an, China}

\title{Rethinking Audio-visual Synchronization for Active Speaker Detection}
\name{%
    Abudukelimu Wuerkaixi$^{\star}$%
    \qquad You Zhang$^{\dagger}$%
    \qquad Zhiyao Duan$^{\dagger}$
    \qquad Changshui Zhang$^{\star}$
}
\address{%
    $^{\star}$ Institute for Artificial Intelligence, Tsinghua University (THUAI), \\%
    State Key Lab of Intelligent Technologies and Systems, \\
    Beijing National Research Center for Information Science and Technology (BNRist), \\
    Department of Automation, Tsinghua University, Beijing, P.R.China \\
    $^{\dagger}$ Department of Electrical and Computer Engineering, University of Rochester, Rochester, NY, USA%
}

\begin{document}

\maketitle

\begin{abstract}
Active speaker detection (ASD) systems are important modules for analyzing multi-talker conversations.
They aim to detect which speakers or none are talking in a visual scene at any given time.
Existing research on ASD does not agree on the definition of active speakers.
We clarify the definition in this work and require synchronization between the audio and visual speaking activities.
This clarification of definition is motivated by our extensive experiments, through which we discover that existing ASD methods fail in modeling the audio-visual synchronization and often classify unsynchronized videos as active speaking.
To address this problem, we propose a cross-modal contrastive learning strategy and apply positional encoding in attention modules for supervised ASD models to leverage the synchronization cue.
Experimental results suggest that our model can successfully detect unsynchronized speaking as not speaking, addressing the limitation of current models~\footnote{The code and demo are available at \url{https://github.com/urkax/SyncTalkNet}.}.
\end{abstract}

\section{Introduction}
\label{sec:intro}




Active speaker detection (ASD) is to determine which person or none is speaking in a video at each time instant. ASD can serve as a fundamental frontend for a variety of downstream tasks including speaker recognition~\cite{chung2018voxceleb2}, speaker diarization~\cite{chung2020spot}, and speech separation~\cite{ephrat2018looking}. Besides, ASD 
can also be applied in human-computer interaction~\cite{cutler2000look} by providing information of
when and where the user is speaking.

The definition of active speakers, however, is not very consistent in the literature. Some works require synchronized speaking signals from the same person on both the audio and visual modalities~\cite{kim2021look}, while others allow non-synchronized speaking signals from different but related persons (e.g., dubbed movies)~\cite{roth2020ava}.
In both cases, speaking activities are required to appear in both audio and visual modalities. 

In this paper, we consider audio-visual synchronization as an essential requirement for ASD
, similar to the definition in the Active Speakers in the Wild (ASW) dataset~\cite{kim2021look}.
We argue that requiring audio-visual synchronization is more reasonable than not requiring it in practice.
This is because it can be ambiguous to define the relevance between visual speaking activities and audio speaking activities. 
Example scenarios include movie dubbing in translated movies, translation voice-over in news video interviews, and narration voice-over in documentaries. Which person should be considered speaking in the video, the person shown in the visual scene, the unseen narrator, or both?
What degree of relevance should be used in the definition of active speakers? In contrast, it is clear whether audio and visual speaking signals are synchronized or not.

\begin{figure}[t]
  \centering
  \includegraphics[width=1.0\linewidth]{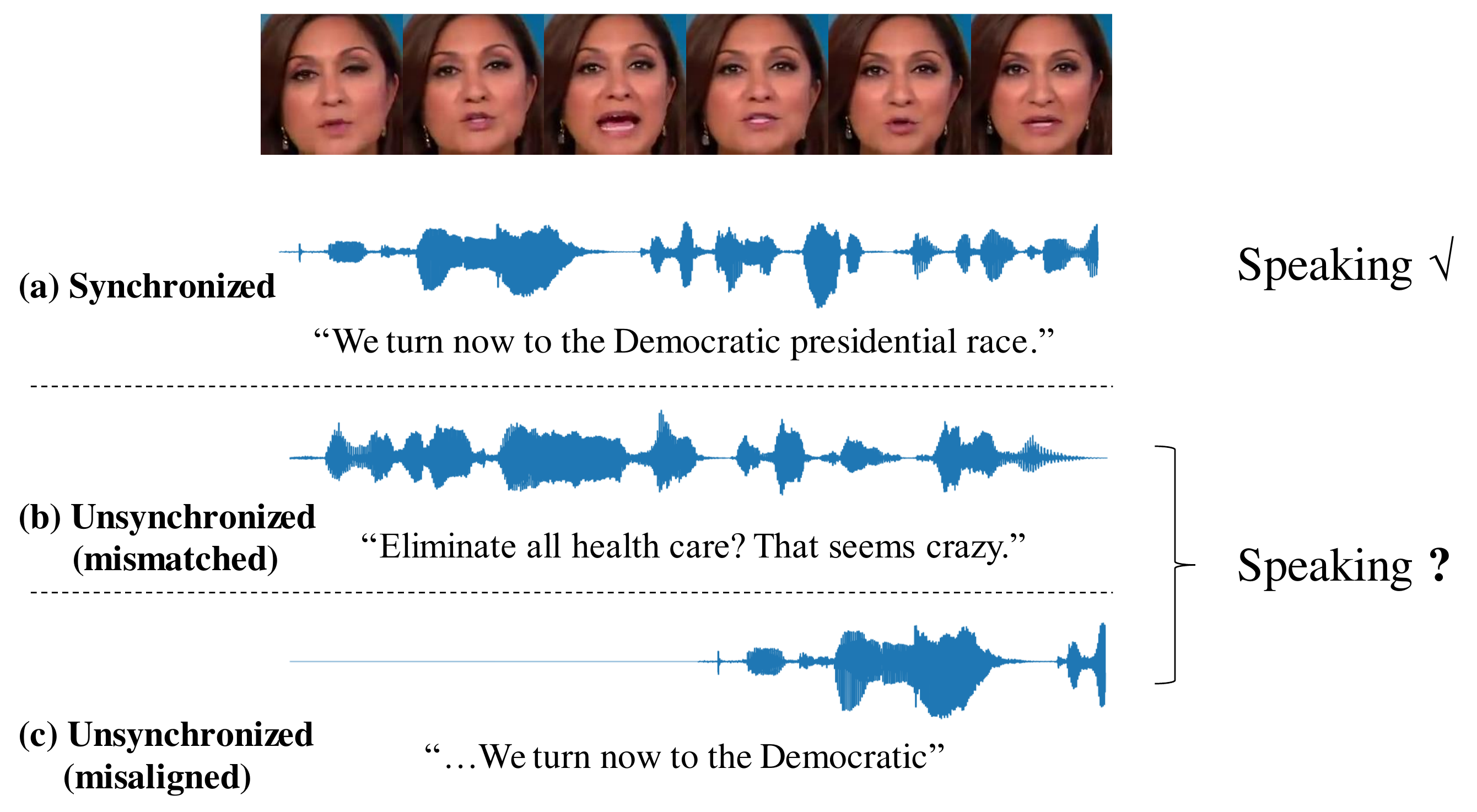}
  \caption{Illustration of a synchronized case (a) and two unsynchronized cases (b, c) for a given face track in ASD. 
  Unsynchronized cases are always considered inactive cases in this work, even if both modalities demonstrate speaking activities.
  }
  \label{fig:unsync_illu}
\end{figure}

Following this definition, it is important to investigate the synchronization issue between audio and visual speaking activities. The investigation of ASD typically starts by tracking faces in the video~\cite{roth2020ava}. A \textit{face track} is defined as a consecutive sequence of faces of a person. For each face track, the audio signal within the onset and offset of the face track is defined as the \textit{audio track}.
If both the face track and the corresponding audio track contain some speaking activities, we can investigate if such signals are synchronized, as illustrated in Figure ~\ref{fig:unsync_illu}. If not synchronized, there are generally two cases:
(1) Mismatch, the audio and the face track do not correspond to the same speaking content;
(2) Misalignment, the content and identity are the same for audio and face tracks, but one modality is delayed.
Perceptual studies found that such delay becomes detectable by ordinary people if it is greater than 125 ms for audio delay and 45 ms for visual delay
~\cite{advanced2003atsc}. 

As reviewed in Section~\ref{sec: rel_work}, existing ASD models do leverage both the audio and visual modalities, however, they do not explicitly model the synchronization between them. It is, therefore, possible that these methods do not consider the audio-visual synchronization cue in ASD. If this is the case, then they would fail on unsynchronized videos, and using such models might be harmful to downstream tasks.

In this work,
we clarify the definition of active speakers, based on which we conduct a case study 
in Section~\ref{sec:case}. We demonstrate that existing supervised ASD models tend to make false-positive predictions on unsynchronized videos.
We believe that such failures are primarily due to two reasons: One is that these models do not explicitly model the synchronization cue. The other is the lack of unsynchronized data in training.
To address these problems, we propose to use cross-modal contrastive learning which is compatible with any supervised ASD method.
We also apply positional encoding in an attention module when fusing audio and visual embeddings, which makes it possible for ASD models to temporally align the two modalities.
Experimental results show that our proposed method performs better than existing supervised and self-supervised models on synthesized unsynchronized videos along with natural videos.
    
    

\section{Related work}
\label{sec: rel_work}
\textbf{Active Speaker Detection}. Researchers have been investigating the task of detecting a talking person in~\cite{cutler2000look,6958874} by taking advantage of audio-visual correlation. With a recent large-scale benchmark dataset proposed in~\cite{roth2020ava} accompanying the atomic visual action (AVA) challenges, more attention has been paid to this task. Most work is based on a multimodal baseline framework that first extracts the audio and visual embeddings from single modalities and then classifies speaking or not speaking segments based on the fused information. Some works focus on improving the modality encoding method~\cite{chung2019naver}, 
some focus on the fusion method~\cite{pouthier21_interspeech, tao2021someone}, and some leverage the context information~\cite{zhang2021unicon, Kopuklu_2021_ICCV}. 
All of the abovementioned methods do not explicitly model audio-visual synchronization. Our work aims to investigate the limitation of such methods by constructing unsynchronization test cases that may fail the current models and then propose methods to address this challenge.

\textbf{Audio-Visual Synchronization}. The audio-visual synchronization in speaking activities has been studied in various contexts.
Owens and Efros~\cite{owens2018audio} proposed to learn audio and visual representations using audio-visual synchronization cues in a self-supervised way.
Chung and Zisserman~\cite{chung2016out} proposed SyncNet to detect the synchronization between lip movement and speech.
There is extensive research based on SyncNet.
Kim et al.~\cite{kim2021look} proposed self-supervised learning with SyncNet for ASD and achieved promising performance.
Ding et al.~\cite{ding2020self} introduced dynamic triplet loss and multinomial loss for self-supervised audio-visual synchronization learning.
Chen et al.~\cite{chen2021audio} studied synchronization in an in-the-wild setting. Cross-modal synchronization modeling has also been studied in audio-visual speech separation~\cite{lee2021looking}, speech-driven talking face generation~\cite{eskimez2021speech}, and lip to speech synthesis~\cite{kim2021lip}.
Our work systematically studies the synchronization modeling in the context of audio-visual ASD.

\section{Case Study}
\label{sec:case}

\begin{figure}[t]
  \centering
  \includegraphics[width=1.0\linewidth]{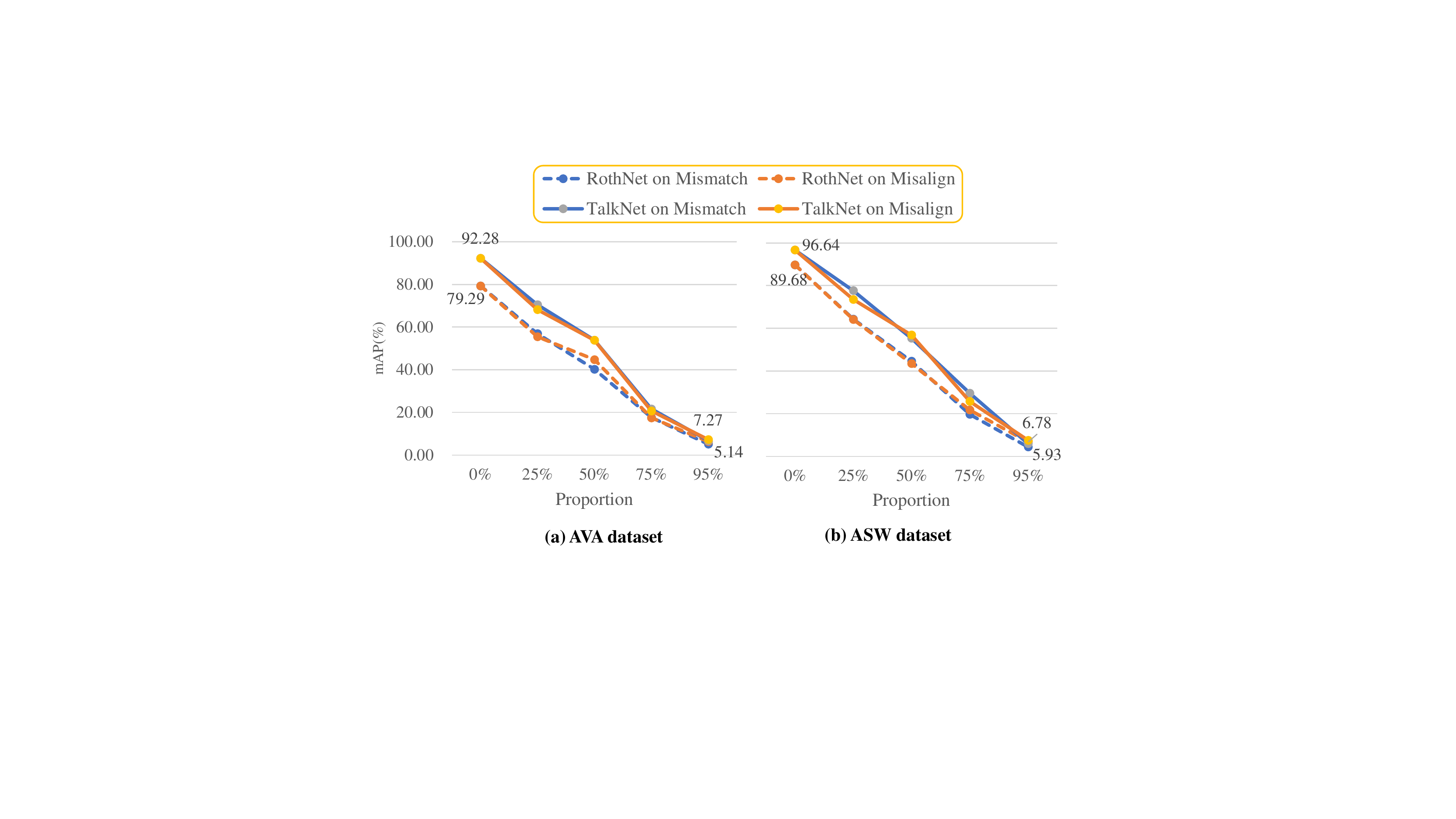}
  \caption{The mean average precision (mAP) of RothNet and TalkNet on retrieving active speaker face frames on the augmented test sets for both AVA and ASW datasets. There are two kinds of augmented test sets for each dataset: augmented with mismatched videos and with misaligned videos. For each kind, we vary the proportion of unsynchronized videos in the test sets, shown as the horizontal axes. 
  }
  \label{fig:casestudy_unsync_test}
\end{figure}

We aim to answer two research questions by studying the performance of current supervised ASD models on unsynchronized audio-visual data. 1) \textit{Can current models correctly label unsynchronized videos as ``Not Speaking''?} 2) \textit{What do current models really learn?}

To the best of our knowledge, current supervised ASD methods do not explicitly model audio-visual synchronization. Also, there are few unsynchronized videos that are correctly labeled in ASD datasets. Therefore, we suspect that current models may not be able to detect unsynchronized videos as ``Not Speaking''. 
With our definition for unsynchronization in Section~\ref{sec:intro}, we augment public benchmarks by synthesizing the audio-visual mismatched and misaligned videos from the original videos to test the existing models.

We use ``RothNet'' to refer to the multimodal framework proposed in~\cite{roth2020ava} as it is a good representative of predominant methods. 
As a common ASD framework, RothNet takes 11 consecutive face frames and their corresponding audio as input, extracts audio and visual features respectively, and concatenates them for further classification.
We also choose the TalkNet model~\cite{tao2021someone} as it achieves the SoTA performance.

\begin{table}[t]
  \centering
  \caption{The audio silence and visual masking experiment on the AVA dataset. The mean average precision (mAP) is displayed.}
    \begin{tabular}{c|cc}
    \hline
          & RothNet & TalkNet \\
    \hline
        AVA   & 79.29\% & 92.28\% \\
    \hline
    Audio silence & 33.61\% & 27.50\% \\
    Visual masking & 39.87\% & 55.20\% \\
    \hline

    \end{tabular}%
  \label{tab:casestudy_slience_mask}%
\end{table}%

\textbf{AVA-ActiveSpeaker (AVA)}~\cite{roth2020ava}.
The AVA dataset is derived from 188 movies around the world.
Each movie is annotated from 15 minutes to 30 minutes with face bounding boxes, entities, and speaking labels. 
And they are human-annotated with ASD labels at the frame level.
There are some dubbed movies in the dataset, and speakers that were active in the original videos are also labeled as active in the dubbed versions. 
However, according to our definition, they should be considered inactive speakers in the videos, given the unsynchronization between the audio and visual modalities.

\textbf{Active Speakers in the Wild (ASW)}~\cite{kim2021look}. The ASW dataset consists of 30.9 hours of face tracks from 212 YouTube videos, including debates, press conferences, talk shows, etc. The length of face tracks ranges from 0.2 to 233.0 seconds. There is no dubbed video in the ASW dataset.

\subsection{Unsynchronization test by augmentation}
\label{sec:case_unsyn}
We propose a mechanism to create unsynchronized video segments (mismatched and misaligned) from original test videos. We then sample from both the unsynchronized videos and the original ones to create augmented test sets where unsynchronized videos take different proportions.
These test sets have the same total number of videos.
We apply such augmentation to both the AVA validation set and the ASW test set.

Specifically, mismatched video segments are created by 
randomly swapping the audio of speaking segments of the original videos.
For each face track, we replace the audio of each speaking segment with another random speaking segment from different videos.
These mismatched speaking segments show lip movements in the video and speaking voices in the audio, but these activities do not match. Their ASD labels are set as negatives. 
For the not-speaking segments in the original videos, they stay the same.




Misaligned video segments are created by shifting the original audio of speaking segments in time. Specifically, we randomly shift the speaking segment's original audio to the left or the right by a time shift greater than 125 ms, which is the human detectable threshold of any delay.
Besides, the shift is circular, i.e., the audio signals shifted beyond one end boundaries are circulated back to the other end.
In this way, the activities within the video segment are preserved, and only the synchroniatinon is destroyed.

The performances of RothNet and TalkNet on the augmented sets with five different proportions of unsynchronized videos are displayed in Fig.~\ref{fig:casestudy_unsync_test}.
As the curves show, 
the mAPs
of the two models degrade rapidly when the amount of unsynchronized data increases. This shows that both ASD models do not properly model audio-visual synchronization. 
We hypothesize that they rely on each individual modality's features and some basic audio-visual correlations to classify videos, but ignore the synchronization cue.



\subsection{Understanding what existing ASD models learn}
\label{sec:case_understand}

To verify that every single modality contributes to the predictions, we remove key information from audio and visual tracks. 
We silence the audio tracks or mask the bottom 30\% of visual frames of each face track with zero to cover the lips in the AVA test set.
As shown in Table~\ref{tab:casestudy_slience_mask}, both models deteriorate dramatically in both cases.
This shows that both models do use voice activity and lip movement information for ASD.


Then, we train a voice activity detection (VAD) model and a lip movement detection model modified from the audio and visual frontends of TalkNet.
The probability of speaking is calculated as the product of the probabilities predicted by the two models.
The mAP of such a combined model in the AVA val set is 90.72\%, which is close to that of TalkNet, indicating that using only a VAD and a lip movement detection model is able to perform comparably with the SoTA ASD models.

\section{Method}

As discussed in Section~\ref{sec:case}, current supervised ASD models fail in unsynchronized cases.
To address the issue, we propose cross-modal contrastive learning, which is compatible with any supervised ASD models.
We apply our method to the SoTA model TalkNet~\cite{tao2021someone} and the predominant baseline model RothNet~\cite{roth2020ava} as two examples.
Apart from cross-modal contrastive learning, we also apply positional encoding in the attention module when fusing audio and visual embeddings.

\begin{figure}[t]
  \centering
  \includegraphics[width=0.9\linewidth]{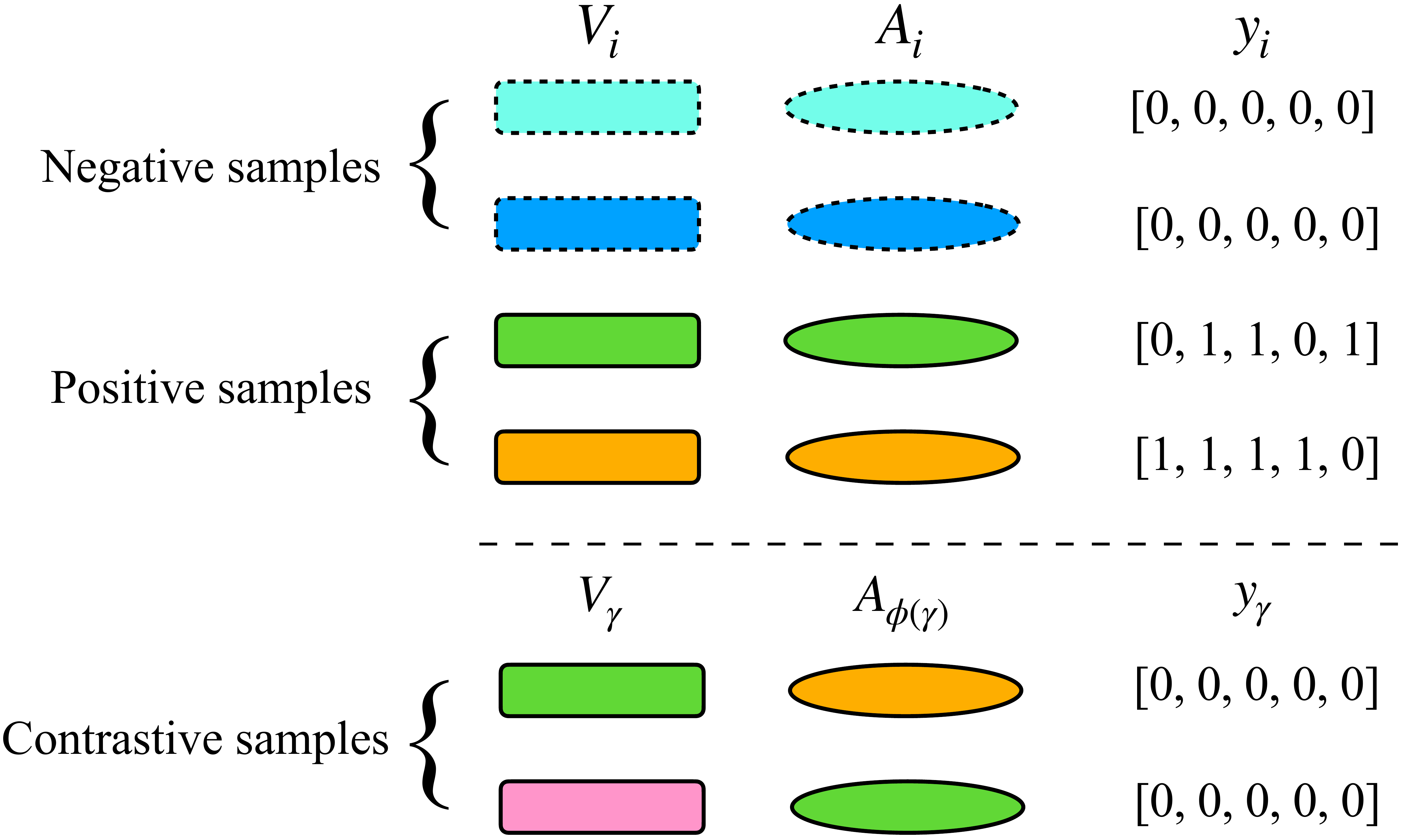}
  \caption{Illustration of the batch formulation in cross-modal contrastive learning. Samples above the dashed line are the features of the original batch, and those below are the artificially unsynchronized samples that are mismatched embeddings. Colors represent embeddings from different samples, shapes denote modalities, and dashed or solid borders denote the ASD labels in the original batch. 
  Better viewed in color.}
  \label{fig:cmcl_illu}
\end{figure}


We elaborate on the architecture and training of Sync-TalkNet
and Sync-RothNet 
in this section, but we believe that our proposed method can be used for any supervised models to encourage synchronization modeling.
The TalkNet and RothNet combined with our method are named as \textit{Sync-TalkNet} and \textit{Sync-RothNet}.

\subsection{Cross-modal contrastive learning}
To alleviate the lack of unsynchronized data in the training dataset, we augment the features in the embedding space to enforce contrastive learning.
Given a batch of embeddings $({V_i, A_i})$ extracted from the frontends  and their corresponding ASD labels $y_i$, where $y_i \in R^T $ and $y_i^t \in \{0, 1\}$, we define $y_i^t = 1$ if the speaker is actively speaking at the given time $t$, otherwise $y_i^t = 0$.
For contrastive learning, as illustrated in Fig.~\ref{fig:cmcl_illu}, we augment a mini-batch of original samples $({V_i, A_i, y_i})$. We define positive samples as the ASD label $y_i$ contains at least one $y_i^t = 1$, and we define such positive samples set as $\Gamma$.
Among the positive samples, we create additional negative sample by randomly exchanging the audio embeddings $A_\gamma$ of a face tracks with audio embeddings of another positive sample $A_{\phi(\gamma)}$, where $\gamma$ and $\phi(\gamma)$ are indexes of two randomly selected face tracks from $\Gamma$, and $\phi(\gamma)$ is different from $\gamma$.
Mathematically, the additional contrastive samples are $(V_\gamma, A_{\phi(\gamma)}, y_\gamma)$, where $\gamma, \phi (\gamma) \in \Gamma$ and $\phi (\gamma) \neq \gamma$.
In the unsynchronized contrastive samples, the visual  face track $\gamma$ is hardly synchronized with the audio from face track $\phi(\gamma)$.
The loss of the training is composed of binary cross-entropy (BCE) loss from the original samples and the loss from contrastive samples, $\mathcal{L} = \mathrm{BCEloss}(f_b(V_i, A_i), y_i) + \beta\cdot \mathrm{BCEloss}(f_b(V_\gamma, A_\phi(\gamma)), \textbf{0})$,
where
$\beta$ controls the balance between the two losses and is set to 1 in our experiments.
The number of frames in a mini-batch is set as 2500, while the number of face tracks varies as in \cite{tao2021someone}.

Although the augmented training set only contains mismatched videos as negative samples, 
as demonstrated in Section~\ref{sec:exp_performance}, such trained models are able to handle not only mismatched videos but also misaligned videos during inference.

\begin{figure}[t]
  \centering
  \includegraphics[width=1\linewidth]{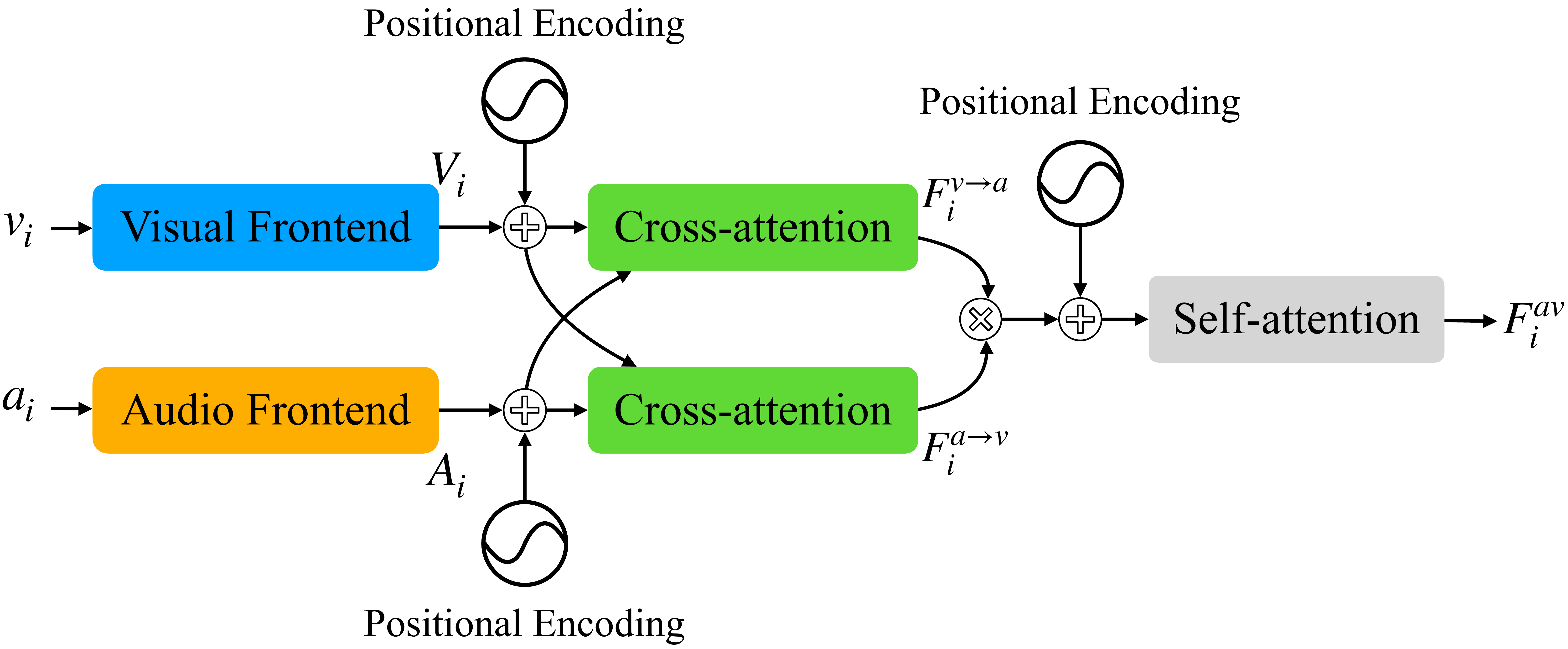}
  \caption{Model architecture of our proposed Sync-TalkNet. 
  }
  \label{fig:pe_illu}
\end{figure}

\subsection{Model architecture: Sync-TalkNet}
\label{sec:model_arch}

The frontend of TalkNet consists of two streams.
The visual frontend takes the $T$ RGB face frames $v_i\in R^{T\times3\times H_v\times W_v}$ of face track $i$ as input, where $H_v$ and $W_v$ are the height and width of the face frames, respectively.
Then the visual embeddings $V_i\in R^{T\times d}$ are computed through a 3D convolutional layer, a ResNet block~, \cite{he2016deep} and a temporal convolutional block.
The audio frontend ingests the Mel-frequency cepstral coefficient (MFCC) vectors $a_i\in R^{4T\times H_a}$ computed from the corresponding audio signals of face track $i$. The audio frame rate is 4 times the 25 FPS video frame rate, and $H_a$ is the dimension of the MFCC vectors.
The audio embeddings $A_i\in R^{T\times d}$ are computed by ResNet34~\cite{he2016deep}.

\begin{table*}[t]
  \centering
  \caption{The mean average precision (mAP) of the five models on the augmented sets and original sets.
  }
  \resizebox{160mm}{!}{
    \begin{tabular}{c|cccc|cc}
    \hline
          & AVA-mismatch & AVA-misalign & ASW-mismatch & ASW-misalign & AVA val set & ASW test set \\
    \hline
    RothNet~\cite{roth2020ava} & 40.24\% & 43.58\% & 44.55\% & 41.17\% & 79.29\% & 89.68\% \\
    TalkNet~\cite{tao2021someone} & 53.91\% & 53.97\% & 55.31\% & 54.79\% & \textbf{92.28\%} & 96.64\% \\
    SyncNet~\cite{chung2016out} & 71.27\% & 61.49\% & 87.21\% & 67.18\% & 82.15\% & 92.40\% \\
    \hline
    \textbf{Sync-RothNet} & 72.14\% & 71.87\% & 79.62\% & 82.37\% & 77.69\% & 89.86\% \\
    \textbf{Sync-TalkNet} & \textbf{79.31\%} & \textbf{74.91\%} & \textbf{87.39\%} & \textbf{89.15\%} & 89.81\%   & \textbf{97.40\%} \\
    \hline
    \end{tabular}%
    }
  \label{tab:exp_performance}%
\end{table*}%

The backend makes predictions of speaking probability in every video frame from the visual and audio embeddings through several attention modules $\mathbf{p}_i=f_b(V_i, A_i)\in R^T$. 
We denote the attention module as $\mathrm{Attention}$:
\vspace{-5pt}
\begin{equation}
    \begin{aligned}
       \mathrm{Attention}(X, Y)&={\rm softmax}(\frac{\textit{Query}(\hat{Y})\textit{Key}(\hat{X})^T}{\sqrt{d}})\textit{Value}(\hat{X}),\\
       \text{where  }&\hat{X}=X+\textit{PE},\text{ }\hat{Y}=Y+\textit{PE}.
    \end{aligned}
\end{equation}
$X$ and $Y$ are encodings from modalities, either audio, visual, or audio-visual.  $\textit{Query}$, $\textit{Key}$, and $\textit{Value}$ are linear layers.
The $\textit{PE}$ is the positional encoding.
Firstly, cross-modal features are computed with cross-attention module, where $F^{a\rightarrow v}_i=\mathrm{Attention}(A_i, V_i)$ and $F^{v\rightarrow a}_i=\mathrm{Attention}(V_i, A_i)$.
Then the $F^{a\rightarrow v}_i$ and $F^{v\rightarrow a}_i$ are concatenated and passed through the self-attention layer $F^{av}_i=\mathrm{Attention}(F^{a\rightarrow v}_i\otimes F^{v\rightarrow a}_i, F^{a\rightarrow v}_i\otimes F^{v\rightarrow a}_i)$.
Lastly, the binary classifier obtains the output from the $F^{av}_i$.

\textbf{Positional Encoding (PE).} Without the positional encoding, the cross-attention layer is permutation-invariant for the inputs, which makes it difficult for the model to learn the synchronization between visual and audio.
Therefore, we add the sinusoidal positional encoding~\cite{vaswani2017attention} $\textit{PE} \in R^{T\times d}$ to the inputs of the cross-attention and self-attention layer,



Besides TalkNet, we also apply our method to RothNet~\cite{roth2020ava}, as described in Section~\ref{sec:case}. RothNet makes predictions for one central frame at each time. We apply cross-modal contrastive learning to train Sync-RothNet.


\section{Experiments}
\subsection{Performance of the proposed method}
\label{sec:exp_performance}

\begin{table}[tbp]
  \centering
  \caption{
  The three highest and three lowest TPRs on the AVA val set of the proposed method.
  }
  \resizebox{90mm}{!}{
    \begin{tabular}{c|ccc}
    \hline
    \hline
    Video ID & \texttt{053oq2xB3oU} & \texttt{P60OxWahxBQ} & \texttt{BCiuXAuCKAU} \\
    \hline
    TPR & 2.14\% & 8.17\% & 9.34\% \\
    \hline
    \hline
    Video ID & \texttt{ax3q-RkVIt4} & \texttt{j5jmjhGBW44} & \texttt{4ZpjKfu6Cl8} \\
    \hline
    TPR & 59.71\% & 61.37\% & 62.67\% \\
    \hline
    \hline
\end{tabular}
    }
  \label{tab:exp_dubbed}%
\end{table}%

\textbf{Unsynchronization test.} To evaluate the performance of the models in unsynchronized cases, we utilize the augmented test sets as described in Section~\ref{sec:case_unsyn}. Using 50\% synthesized data (either mismatched or misaligned) and 50\% original data, we curate AVA-mismatch, AVA-misalign, ASW-mismatch, ASW-misalign sets.
We test Sync-TalkNet and Sync-RothNet on the augmented and original test sets.
Table~\ref{tab:exp_performance} displays the results.
Sync-TalkNet and Sync-RothNet significantly outperform the two supervised baseline models RothNet and TalkNet in the augmented test sets.
In particular, Sync-TalkNet achieves better results than the SoTA model TalkNet on an original dataset: the ASW test set.
The mAP of Sync-TalkNet is slightly lower than TalkNet on the AVA val set.
It is reasonable because the AVA dataset contains dubbed movies in both the training and the validation sets;
The unsynchronized speaking segments are labeled as positive in this dataset, damaging the performance of Sync-TalkNet, which leverages synchronization cues.
Cross-modal contrastive learning loss is similar to the augmentation of mismatched data, while Sync-TalkNet and Sync-RothNet also work well on misaligned data.
This verifies the ability of synchronization modeling in our proposed method.

Sync-TalkNet and Sync-RothNet are trained in a supervised manner. The self-supervised ASD models such as SyncNet~\cite{chung2016out} is trained by distinguishing matched and mismatched audio-visual pairs. Thus SyncNet is also able to detect audio-visual synchronization as shown in Table~\ref{tab:exp_performance}, achieving comparable results to our model in the unsynchronized mismatched test sets. However, our proposed models as supervised models perform better than SyncNet on the original datasets and unsynchronized misaligned test sets.
Generally speaking, supervised models gain better capacity for usual ASD, while self-supervised models better leverage the synchronization cue, since self-supervised models are trained through synchronization classification.
Our proposed method is able to leverage both advantages, achieving excellent performance on both the original datasets and unsynchronization augmented datasets. 


\textbf{Narrated videos detection.} 
According to our definition, any face track in dubbed movies should not be regarded as speaking. However, there are some dubbed movies in the AVA dataset and they are labeled as speaking which misleads the training and evaluation of ASD models with synchronization cues.
We use Sync-TalkNet trained on the ASW dataset to filter out the dubbed movies in the AVA validation set.
We calculate the true positive rate (TPR), the ratio of the positively predicted frames to the positively labeled frames. Lower TPR indicates the more likely the video is from a dubbed movie.
In the dubbed movies of the AVA val set, the positively labeled frames are from the unsynchronized speaking segments.
Out of the 33 videos, we respectively display the videos with the three highest TPR and the three lowest TPR in Table~\ref{tab:exp_dubbed}.
Through manual checking, we verified that the three videos with the lowest TPR are dubbed movies, and the three with the highest TPR are not dubbed movies.
Although the samples are limited, this result does suggest that Sync-TalkNet might be able to be used to detect unsynchronization in narrated videos.




\begin{table}[tbp]
  \centering
  \caption{Ablation study of cross-model contrastive learning and positional encoding on the ASW augmented test sets.}
  \resizebox{80mm}{!}{
    \begin{tabular}{c|cc}
    \hline
          & ASW-mismatch & ASW-misalign \\
    \hline
    \textbf{Sync-TalkNet} & \textbf{87.39\%} & \textbf{89.15\%} \\
    \hline
    w/o contr. learning & 56.22\% & 54.71\% \\
    w/o both PE & 84.06\% & 86.24\% \\
    w/o cross-attention PE & 84.49\% & 84.37\% \\
    w/o self-attention PE & 85.99\% & 87.37\% \\
    \hline
    \end{tabular}%
    }
  \label{tab:exp_ablation_part}%
\end{table}%

\subsection{Ablation study}

To model synchronization, we apply positional encoding and cross-modal contrastive learning.
We conduct an ablation study of Sync-TalkNet for these two modules as shown in Table~\ref{tab:exp_ablation_part}.
According to the results, cross-modal contrastive learning plays an important role in learning synchronization.
Removing the positional encoding, the performance degrades but does not degrade catastrophically. We believe that with the guidance of contrastive learning, the model learns the weak timeline information through the zero-padding of convolutional layers in the frontend~\cite{kayhan2020translation}. However, our model with the positional encoding 
still performs better.

As introduced in Section~\ref{sec:model_arch}, there are cross-attention and self-attention modules in the model architecture.
Sync-TalkNet applies positional encoding on both the attention modules. We study the effect of disposing of one of the two positional encodings. As the results in Table~\ref{tab:exp_ablation_part}, applying positional encoding on both the attention module helps Sync-TalkNet better perceive the timeline information.


\section{Conclusion}
In this paper, we argued that audio-visual synchronization is an important cue for active speaker detection (ASD), and clarified the definition of active speakers to explicitly require audio-visual synchronization. This was motivated by our experiments which showed that state-of-the-art supervised ASD models neglect this synchronization in their decision-making. Not modeling synchronization would make them unreliable for many downstream tasks and real-life applications.
We then proposed to apply cross-modal contrastive learning and positional encoding in the attention modules to supervised ASD models. Our method outperforms supervised and self-supervised models in artificially unsynchronized cases and regular benchmarks.
For future work, we plan to broaden the definition of active speakers to speaking activities shown in either modality; This would include the audio-visual synchronized cases in this paper, as well as speaking activities that are invisible or inaudible.




\small

\bibliographystyle{IEEEbib}
\bibliography{strings,refs}

\end{document}



\maketitle

\begin{figure*}[h]
  \centering
  \includegraphics[width=1\linewidth]{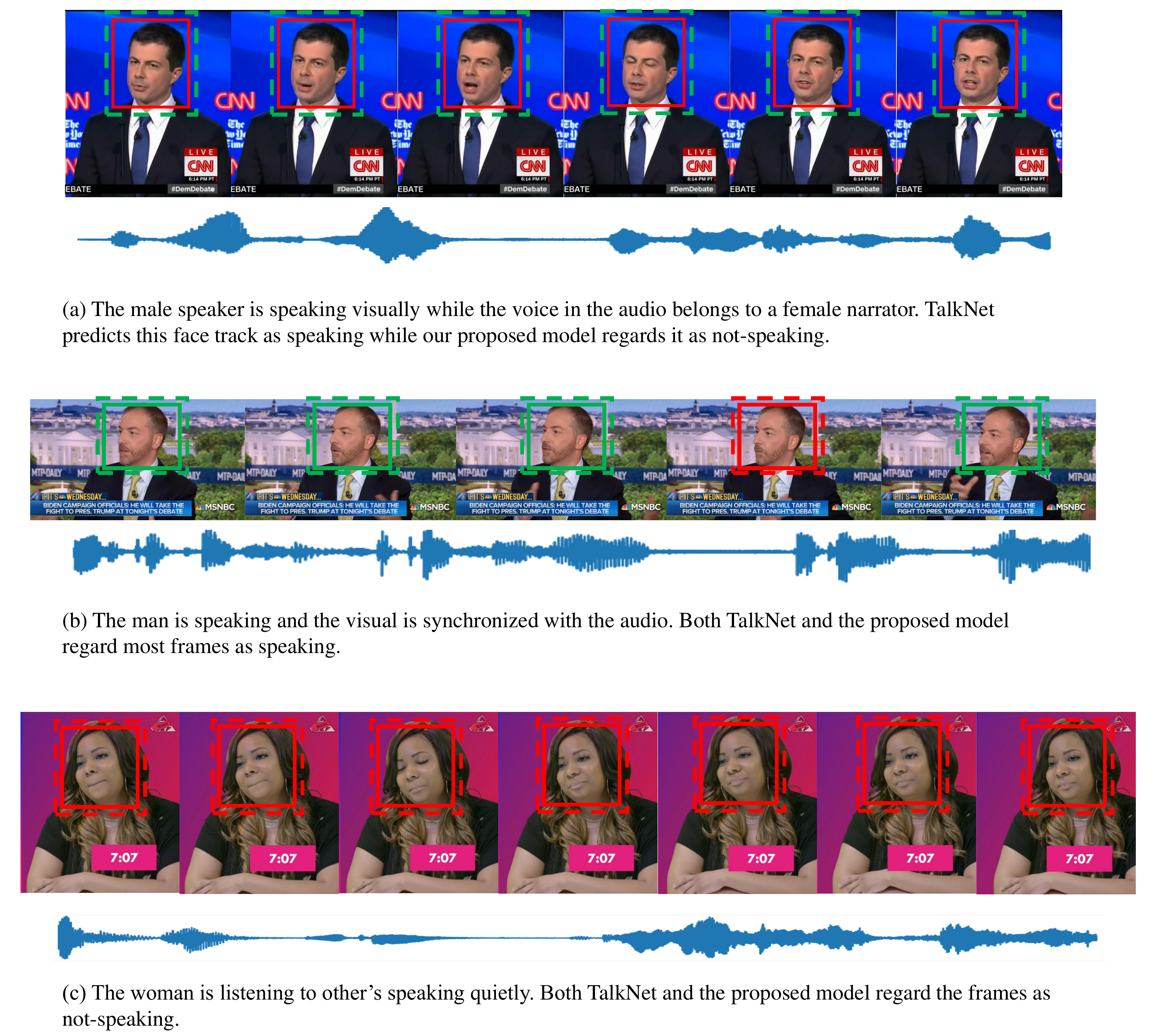}
  \caption{Active speaker detection results of three video segments from the ASW dataset. The dashed rectangles represent the predictions of TalkNet, and the solid rectangles represent the predictions of Sync-TalkNet. The color of green and red respectively indicate positive and negative predictions. The example (a) is a special case where the visual and audio signal can be regarded as speaking independently (the mouth is moving in the visual and there is a voice speaking \zd{from another person} in the audio). In this case, TalkNet \zd{regards} the segment as speaking \zd{which is incorrect according to our definition}. Sync-TalkNet can handle this case. The examples (b) and (c) are regular speaking and not-speaking circumstances, Sync-TalkNet and TalkNet both make correct predictions.
  }
  \label{fig:examples}
\end{figure*}

\subsection{Examples}
Unsynchronization is not uncommon in real-life applications. For example, in Figure~\ref{fig:examples} (a), the male speaker is speaking in the video with his voice muted, and a female narrator is speaking out of the sight. As previous models detect active speaker based on mouth movement and voice activity detection, ignoring the synchronization, false positives are produced. \zd{In contrast, our proposed model makes a not-speaking prediction as it requires audio-visual} synchronization.

\subsection{Sync-RothNet}
As described in Section 3, RothNet is a two-steam model.
The audio-visual embeddings are respectively extracted using deep neural networks.
Then the two embeddings are concatenated and passed through a binary classifier to obtain the speaking probability prediction.
RothNet takes 11 consecutive face frames and the corresponding audio as input, and makes prediction for the central frame.
There is no attention module in RothNet among different time steps, so the positional encoding is not included.
We apply cross-modal contrastive learning to train Sync-RothNet.